\documentclass[12pt]{article}
\usepackage{graphics}
\begin{document}
\begin{center}
{\large Calculation of Spin Observables for Proton-Proton Elastic Scattering in the Bethe-Salpeter Equation}
\end{center}
\begin{center}
{Susumu Kinpara}
\end{center}
\begin{center}
{\it National Institute of Radiological Sciences, Chiba 263-8555, Japan}
\end{center}
\begin{abstract}
Bethe-Salpeter equation is applied to $p$-$p$ elastic scattering. 
The observables of spin are calculated in the framework of the M matrix using the two-body interaction potential.
The parameter of the pseudovector coupling constant is adjusted so as to reproduce the spin singlet part.
It is shown that the spin rotation $R(\theta)$ and $A(\theta)$ are improved by
the resonance effect for ${}^{\rm 1}S_{\rm 0}$.
\end{abstract}
\hspace*{4.mm}
Elucidation of nucleon-nucleon two-body system in terms of the meson exchange model 
is the main purpose in the present study.
The nuclear force between nucleons is mediated by the various mesons and their creation and annihilation are 
described by the method of the field theory. 
To relate the phenomenologically accepted ideas of the meson theory with the dynamical mechanism of two-body system
the Bethe-Salpeter (BS) equation is indispensable and the investigation for a long time has revealed the applicability
to various phenomena on the composite system.
\\\hspace*{4.mm}
Recently our calculations for nucleon-nucleon elastic scattering have shown that the BS equation
is suitable not only for bound state that is deuteron also for the scattering state$\cite{Kinpara}$.
In the study the lowest-order calculation known as the Born approximation is corrected by the higher-order calculation
and the results of the differential cross section and the polarization are improved to make possible to compare with
the experimental data quantitatively.
We proceed to calculations for various observables of spin in the elastic scattering as described later on.
\\\hspace*{4.mm}
One of the characteristics in the present formulation by the BS equation is the inverse square potential.
The virtual creation and annihilation process due to the mesons is represented by the Feynman propagator
and for example the form of the $\sigma$ meson is expressed 
by using the modified Bessel function of the first order as $\sim K_{\rm 1}(r)/r$ in the coordinate space.
Expanding the BS amplitude by the set of the Gamma matrix and imposing the auxiliary condition the resulting equation 
appears to be equivalent to the Schr$\rm{\ddot o}$dinger eigenvalue equation 
with the potential terms as above on the various mesons.
So, the leading inverse square part ($\sim {\rm 1}/r^{\rm 2}$) is stronger than the screened Coulomb potential 
(Yukawa potential) at the short-range region and expected to have an effect on the two-body system.
\\\hspace*{4.mm}
Another important character of the BS equation is that the BS amplitude consists of 16 expansion coefficients
which can act as the wave functions of the relative coordinate between two nucleons.
The role of each component is made clear by studying the property of spin.
The set of three polar-tensor components is assigned the spin triplet ($S$=1)
and therefore suitable for constructing deuteron and the scattering state.
Besides, the spin singlet ($S$=0) state is another object to be considered 
and also possible to understand it within the BS equation.
It is the point at issue in the present formulation to select the most proper component 
among three, that is, the zeroth vector, the zeroth axial-vector and the pseudoscalar ones.
While the zeroth vector component is independent of the other BS components in the simultaneous equations,
the spin function $\it\Gamma_V$ of it is symmetric (${}^t\it\Gamma_V=\it\Gamma_V$) under the exchange 
between two fermions and then set identically equal to zero.
On the remaining two components there is no reason which one is advantageous as the $S$=0 wave
and it is one of the main subjects to determine it by comparing them with the experimental data.
\\\hspace*{4.mm}
Difference between the zeroth axial-vector ($av$) and the pseudoscalar ($ps$) equations for the $S$=0 part is represented 
by the strength of the leading order inverse square potential part 
$V_i(r)=-g_i M^{-1} /r^2+O(r^0)\:$($i=av$ or $ps$) given as
\begin{equation}
g_{av}=\frac{-g_\sigma^2-2 g_\omega^2}{(2 \pi)^2},
\end{equation}
\\\hspace*{4.mm}
\begin{equation}
g_{ps}=\frac{g_\sigma^2-4 g_\omega^2}{(2 \pi)^2}-\frac{(4T-3) f^2{\mit\Lambda}^2}{(2 \pi)^2m_\pi^2},
\end{equation}
in which the isospin $T=0,1$ of two nucleon system and $M$ is the nucleon mass.
For the isoscalar $\sigma$ and $\omega$ mesons contribute dominantly, both $g_{av}$ and $g_{ps}$ are
negative values and thus the interaction of the $S$=0 part may become a repulsive force
at the intermediate energy.
The signs of $g_{av}$ and $g_{ps}$ are significant to construct the proton-proton ($p$-$p$) elastic scattering 
because the virtual ($S,T,L$)=(0,0,1) state is introduced to make the phase shift parameter of the (1,1,1) state
remains a real number$\cite{Kinpara}$.
In the present calculation either $g_{av}$ or $g_{ps}$ dependence of the higher-order term is very large, then, to compare 
these two models we set $g_{av}=g_{ps}$ 
and use the individual values of the coupling constant $g_\omega$ of $\omega$ meson.
It is fortunate that the results of the calculation are not sensitive to $g_\omega$ 
because of the heavier value of the mass ($m_\omega$/$m_\pi={\rm 5}\sim{\rm 6}$).
\\\hspace*{4.mm}
In order to examine the meaning of each component of the BS amplitude 
the spin observables are appropriate to draw some conclusions 
and to find out the two-body interaction between nucleons. 
When one evaluates the spin observables it is formulated in terms of the density matrix$\cite{Hoshizaki}$ 
and the phenomena on the elastic scattering between two nucleons are represented by the quantities on the isospin $T$ as
\begin{equation}
a^T_{\mu\nu\tau\rho}
\equiv\frac{1}{4}{\rm Tr}\,M^T \sigma_\tau^{(1)} \sigma_\rho^{(2)} M^T{}^\dagger \sigma_\mu^{(1)} \sigma_\nu^{(2)}\qquad
(T=0,1),
\end{equation}
where the subscript $\mu$ of $\sigma_\mu^{(1,2)}$ for particle 1 or 2 
is $\mu=0,x,y,z$ and so on ($\sigma_0^{(1,2)}\equiv1$).
Extension of spin to the arbitrary directions is done similarly.
Although the actual calculation of $a^T_{\mu\nu\tau\rho}$ is restricted to which the number of the subscripts except 0
is at most two in the present study, the applications to the higher order quantities are possible.
\\\hspace*{4.mm}
The elastic scattering is resolved by the method of the M matrix which has already been reported 
in our previous study$\cite{Kinpara}$.
In it, to specify each partial wave of the M matrix the nuclear bar phase shift parametrization is used 
with no inclusion of the Coulomb force.
The elements of the K matrix are associated with the standing wave solutions.
In place of the exact solution the approximate form of the wave function under the leading order inverse square part 
of the potential derived from the BS equation is used in which the regularization procedure due to the cut-off
function in the pion exchange interaction is taken into account.
Our procedure to regularize the divergences caused by the inverse fourth power potential 
from the pseudovector coupling interaction
acts only on the Born term, nevertheless, it has been verified that the results of calculation work well.
\\\hspace*{4.mm}
A point to note is that in the present formulation the charge independence is broken.
There are two reasons.
For evaluation of the $p$-$p$ elastic scattering within the ladder approximation 
the pion-nucleon coupling constant $f$ of the pseudovector interaction is reduced 
from the standard value of it to the $\sim1/20$.
Adding that the ${}^3P_2$ state is corrected by the higher-order term using the inverse square part of the potential.
Consequently the isospin $T$=1 part of the M matrix ($M^1$) is different between $p$-$n$ and $p$-$p$
case through the phase shift parameters.
\\\hspace*{4.mm}
Calculation of the M matrix is done in the center of mass system 
at the azimuthal angle $\varphi=0$ for the scattered nucleon.
The elements of the M matrix are represented 
by the composite spin states $|s\!>\equiv|00\!>$ and $|s_z\!>\equiv|1s_z\!>$.
The irreducible representation of $\sigma_\mu^{(1,2)}$ required 
is connected to the direct product representation by the unitary transformation.
The matrix elements between spin singlet and the triplet $M^T_{ss_z}$ and $M^T_{s_zs}$ vanish
as the result of the parity conservation of the interaction for two identical fermions.
The remaining 10 elements are reduced to 
$M^T_{ss}$, $M^T_{11}$, $M^T_{00}$, $M^T_{1-1}$, $M^T_{01}$ and $M^T_{10}$
on account of the space reflection invariance of the M matrix.
Consequently all spin observables 
of two nucleon elastic scattering are prepared as a function of the center of mass angle and the incident energy.
\\\hspace*{4.mm}
The elements of the spin triplet are not independent 
and which are related to each other by the time reversal invariance relation$\cite{Wolfenstein}$.
To apply it in the present calculation the general relation of $M(\theta\,\varphi)$ is used at $\varphi=\pi$.
It is alternatively performed by the unitary transform on $M(\theta\,\varphi)$ to reverse the scattering plane 
before taking the trace with the term which breaks the time reversal invariance at $\varphi=\rm 0$.
Then, it is verified that the condition of the time reversal invariance is satisfied in the numerical calculation.
\\\hspace*{4.mm}
While our calculation with the BS equation is on the basis of the relativistic framework of the theory 
and the Gamma matrix expansion appears to maintain it, the transformation 
between the center of mass and the laboratory system is done within the nonrelativistic kinematics for simplicity.
It is conveniently performed by introducing the triad of the unit vectors ($\hat P$,$\hat K$,$\hat n$)
defined as $\hat P\equiv(\vec{k}+\vec{k'})/\vert\vec{k}+\vec{k'}\vert$, 
$\hat K\equiv(\vec{k'}-\vec{k})/\vert\vec{k'}-\vec{k}\vert$ 
and $\hat n\equiv\hat P\times\hat K=\hat y$, in
which $\vec{k}$ ($\parallel \hat z$) and $\vec{k'}$ are momenta of the incident and the scattered nucleons 
in the center of mass system respectively.
The direction of the final state scattered nucleon ($\hat{p}_s$) and that of the recoil nucleon ($\hat{p}_r$) 
in the laboratory system is represented by the unit vectors as $\hat{p}_s=\hat P$ and $\hat{p}_r=-\hat K$.
The side directions of the final state scattered and recoil nucleons 
in the laboratory system are given 
by $\hat n\times\hat{p}_s=\hat K$ and $\hat n\times\hat{p}_r=\hat P$ accordingly.
\\\hspace*{4.mm}
Investigating the difference between the $av$ and $ps$ equations for the $S$=0 part 
and the role on $p$-$p$ elastic scattering 
the spin-correlation parameter $C_{nn}\equiv a^{1}_{yy00}/I_0$ $(I_0\equiv a^{1}_{0000})$ is useful 
because at the center of mass scattering angle $\theta=90^\circ$
the quantity $I_0(1-C_{nn})=\vert M^{1}_{ss} \vert^2/2$ is independent of the $S$=1 of the M matrix
and makes determine the singlet part $M^{1}_{ss}(\theta=90^\circ)$.
\\\hspace*{4.mm}
The quantity $I_0(1-C_{nn})$ of the $p$-$p$ elastic scattering at $\theta=90^\circ$ by the calculation 
is plotted in Fig. 1.
The laboratory energy of the incident proton is 310 MeV.
The experimental value of $C_{nn}$ is given by $\sim$ 0.6 at the 300 $\sim$ 400 MeV laboratory energy$\cite{Beretvas}$
and the related experimental value of $I_0(1-C_{nn})$ is 3.7$\times$(1$-$0.6)$\,\approx\,$1.48 
which is shown by the dotted line.
It is seen that the result of the $av$ equation reproduces the $M^1_{ss}(\theta=90^\circ)$ part 
at the intermediate energy.
The $ps$ equation is about threefold as large as the experimental value and is not likely to reach the experimental area.
\\\hspace*{4.mm}
It is necessary to examine the $f$ dependence of the $S$=1 part of the M matrix too.
For the $\omega$ meson interaction is practically zero in the interaction of the polar-tensor equation
giving the $S$=1 wave, there is no model dependence between the $av$ and the $ps$ models.
The quantity $I_0C_{KP} \equiv a^{1}_{KP00}$ calculated at $\theta=90^\circ$ is plotted in Fig. 2.
The spin correlation parameter $C_{KP}$ by the measurement is seen in the ref. $\cite{Engels}$. 
In order to adapt it to the definition of $C_{KP}$ here the relation $C_{KP}=-C_{-KP}$ is used.
The form of $I_0C_{KP}=(\vert M^{1}_{01} \vert^2-\vert M^{1}_{10} \vert^2)/2$ at $\theta=90^\circ$ indicates
the higher-order correction to the phase shift of the ${}^3P_2$ state may influence the numerical value much.
Comparing with the experimental value at the 400 MeV laboratory energy ($I_0C_{KP}=3.7\times(-0.32)\approx-1.18$)  
the optimum value of $f$ is around 0.05, a little smaller than the value of the $S$=0 case.
Using another result of the measurement$\cite{Engels}$ the experimental $I_0C_{KP}$ value doubles
and the point of intersection approaches the expected region.
\\\hspace*{4.mm}
To investigate the value of $f$ furthermore we take up the differential cross section $I_0$
which contains all the elements of the M matrix and has a basic meaning for the calculation of various spin observables.
In Fig. 3 the $f$ dependence is shown from 0.05 to 0.08 at $\theta=90^\circ$ and the 310 MeV laboratory energy. 
The $av$ equation gives the tentative value of $f$ at $\sim$ 0.065 well.
Due to the overestimate of the $M^1_{ss}(\theta=90^\circ)$ value the result of the $ps$ equation is larger than that of the $av$ equation.
\\\hspace*{4.mm}
Next our interest is the angular dependence of various spin observables by using the value of the parameter $f=0.065$.
The differential cross section of $p$-$p$ elastic scattering at the 310 MeV laboratory energy is plotted in Fig. 4.
The angular dependence is not seen noticeable to the difference between the $ps$ and $av$ equations.
It implies that in the singlet part $M^1_{ss}(\theta)$ the $S$-wave may play a dominant role  
compared with the higher waves even at the intermediate energy.
\\\hspace*{4.mm}
The polarization $P(\theta) \equiv a^{1}_{n000}/I_0$ is suitable to observe the properties besides $\theta=90^\circ$
since experimentally the maximum is observed at around $\theta=30^\circ\sim40^\circ$
and the relation $P(\theta)=-P(\pi-\theta)$ due to the Pauli principle restricts the area under consideration.
In Fig. 5 the polarization as a function of $\theta$ at the 400 MeV laboratory energy 
is shown for the $av$ and $ps$ equations.
The quantity $I_0P(\theta)$ is independent of $M^1_{ss}(\theta)$ and
the underestimate of $P(\theta)$ is connected with the overestimate of $I_0(\theta)$ in the case of the $ps$ equation.
On the other hand the $av$ equation gives the sufficient maximum value
in exchange for appearance of the dip at $\theta\sim40^\circ$ in the differential cross section.
The magnitude and the position of the maximum value is understood by the enhancement of the ${}^3P_2$ state
due to the higher-order correction, that is, 
$P(\theta) \sim P_1(\rm{cos} \theta) \cdot {\it P}_1^1(\rm{cos} \theta)\sim \rm{sin} \rm{2}\theta$ 
giving the position a rather shifted to the right than that of the experimental data$\cite{Besset}$$\cite{Besset2}$.
\\\hspace*{4.mm}
The depolarization $D(\theta) \equiv a^{1}_{y0y0}/I_0$ covers the whole scattering angles 
and many experimental data have been accumulated nowadays$\cite{Besset}$$\cite{Onel}$.
In Fig. 6 the depolarization as a function of $\theta$ at the 400 MeV laboratory energy 
is shown for the $av$ and $ps$ equations.
The experimental $D(\theta)$ values over $\theta=90^\circ$ are plotted 
by using the relation $D(\theta)=K(\pi-\theta)$, in which
the spin transfer $K(\theta)$ is defined as $K(\theta) \equiv a^{1}_{y00y}/I_0$.
At $\theta=30^\circ\sim40^\circ$ the result of $D(\theta)$ overestimates the experimental data as well as
the $av$ case in the polarization.
In the present formulation only the $P$-wave is corrected by the higher-order terms making
the partial wave be strong particularly$\cite{Kinpara}$.
Then, the extension of the method to include the other partial waves may improve the results of the calculation.
\\\hspace*{4.mm}
In addition to the effect of the higher-order terms beyond the Born approximation, resonant properties
are thought to be essential for $p$-$p$ elastic scattering at the intermediate energy.
Particularly observables on the triple-scattering experiments contain $M^1_{ss}(\theta)$ in the numerator 
and the correction may change the angular distributions largely.
The spin rotation parameters $R(\theta)$ and $A(\theta)$ are concerned with rotation of the spin to the side direction
$\hat K$ on the scattered proton initially directed to $\hat x$ and $\hat z$ 
and which are given by $a^{1}_{K0i0}/I_0$ with $i=x$ and $z$ respectively.
The calculations are performed as well as the procedure for $D(\theta)$ 
and compared with the data of the measurements below $\theta=90^\circ$$\cite{Besset}$$\cite{Onel}$.
\\\hspace*{4.mm}
In Fig. 7 the parameter $R(\theta)$ is shown as a function of the center of mass angle $\theta$ from $\theta=0^\circ$ to $90^\circ$ at the 400 MeV laboratory energy.
It is clearly seen that both $av$ and $ps$ equations do not give the results convincingly.
Although the difference between the calculation and the experiment is not small, it leaves room for 
changing the situation by taking into account the effect of the resonance.
\\\hspace*{4.mm}
As the incident proton energy approaches the region of the pion production threshold 
the $p$-$p$ system begins to scatter elastically via the process $p+p\rightarrow\pi^++p+n$ 
primarily in the spin singlet $S$-wave.
Thus, the phase shift parameter $\delta_L$ ($L$=0) is corrected by the resonant term and which is represented by\\
\begin{equation}
\delta_0 \rightarrow \delta_0 - \rm{tan}^{-1}\frac{\it\Gamma}{2(\it E-\it E_{\rm 0})},
\end{equation}
in which $E$ and $E_0$ are the incident energy of proton 
and the resonant energy located in the intermediate energy respectively$\cite{Landau}$.
The intermediate quasi-stationary state ($\pi^++p+n$) has the width $\it\Gamma$ 
associated with the life time $\tau$ as $\it{\Gamma} \sim \rm 1/\tau$.
At present the definite value of $\it\Gamma$ is not required because $E \sim E_{\rm 0}$ 
and so the second term of Eq. (4) gives $\pm \pi/\rm2$ 
which acts on the element of the S matrix $S_{\rm 0} = \exp ({\rm 2} i \delta_{\rm 0})$ to reverse the sign.
\\\hspace*{4.mm}
The resonant state is expected to make effects particularly at $\theta\sim90^\circ$
because only three elements of the M matrix $M^1_{ss}$, $M^1_{01}$ and $M^1_{10}$ remain therein 
and $I_{\rm 0} R(\theta)$ is shown by\\
\begin{equation}
I_0\,R(\theta)=\frac{1}{2}\,\rm{Re}[\,M^1_{\it ss}(\theta)\,M^1_{01}(\theta)^\ast]\;\;\;\;\;\;\;\;\;(\theta=90^\circ).
\end{equation}
The quantity $I_0\,A(\theta)$ is analogous to $I_0\,R(\theta)$ 
and it is given by replacing $M^1_{01}$ with $-M^1_{10}$ in Eq. (5).
\\\hspace*{4.mm}
Taking into account the resonance mentioned above the $av$ equation denoted by $av(\pi/\rm 2)$ nearly 
reverses the sign at $\theta=90^\circ$ in comparison of the calculation without the resonance effect in the $R(\theta)$.
It makes possible to explain the trend of the experimental data from $\theta=10^\circ$ to $50^\circ$ well.
The other resonance, for example, the $P$-wave is hoped to explain the trend at $\theta\geq50^\circ$.
For the calculation does not include the electro-magnetic interaction the sharp edge around $\theta=10^\circ$
arising from the interference is not attained to describe here.
While the approximate relation $R(\theta,\delta_0+\pi/2) \sim -R(\theta,\delta_0)$ is seen in the case of the $av$ equation,
the $ps$ equation does not show the feature which could be attributed to the difference of the phase shift 
of the higher waves ($L\geq\rm 2$) in the spin singlet part.
\\\hspace*{4.mm}
The resonance is applicable to the other spin parameters too.
In Fig. 8 the spin rotation $A(\theta)$ is shown as a function of the center of mass angle $\theta$ from $\theta=0^\circ$ to $90^\circ$ at the 400 MeV laboratory energy.
At $\theta\le20^\circ$ the results of calculation show the trend to decrease along with the experimental data.
Without the effect of the resonance the results do not follow the sudden rising observed in the measurements.
By including the resonance the resultant curves denoted by $av(\pi/\rm 2)$ and $ps(\pi/\rm 2)$ turn to increase 
as far as intersecting the experimental data at $\theta\sim70^\circ$.
The remaining differences over $\theta\sim70^\circ$ are not large in comparison of the case of $R(\theta)$. 
\\\hspace*{4.mm}
The improvement of $R(\theta)$ by the resonance effect makes us apply it also to $D(\theta)$ 
for satisfying the relation $R(\theta)=D(\theta)$ at $\theta=0^\circ$ 
regardless of lack of the electro-magnetic interaction.
It has been verified that the overestimate at $\theta=30^\circ\sim80^\circ$
in the calculation of $D(\theta)$ is remedied by including the resonance effect.
\\\hspace*{4.mm}
In the present study proton-proton elastic scattering is calculated by using the Bethe-Salpeter equation.
The change of the pseudovector coupling constant of pion is inevitable to reproduce the experimental data.
Consequently the structure of the M matrix is modified 
and in fact the charge independence of the two-nucleon system is broken.
The spin parameters in the triple-experiment are understandable if we take account of the resonance caused
by the $\pi^++p+n$ quasi-stationary state.

\newpage
${\bf Figure\;Captions}$\\\\
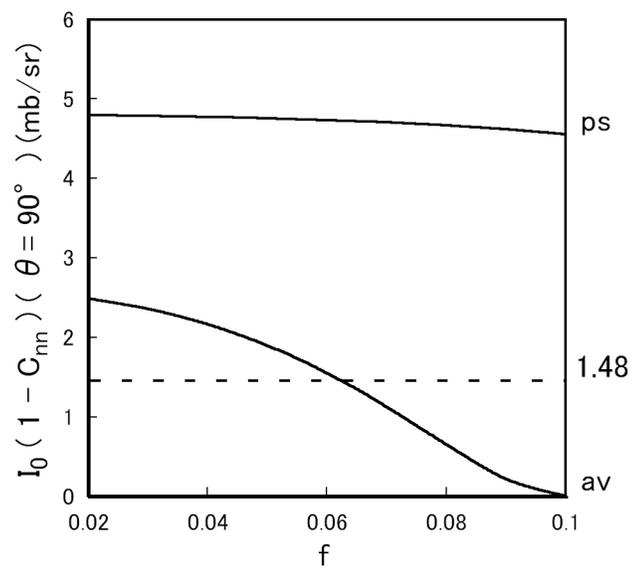
Figure 1: $I_{\rm 0}(1-C_{nn})$ as a function of the pseudovector coupling constant $f$
at the center of mass scattering angle $\theta=90^\circ$ and the laboratory energy of 310 MeV. 
The $av$ and $ps$ denote the axial-vector and the pseudoscalar components for spin singlet part respectively.
The experimental value 1.48 is from $C_{nn}$ data of 330 MeV in ref. $\cite{Beretvas}$ 
and $I_0$ value at 312 MeV and $\theta=30.67^\circ$ in ref. $\cite{Besset}$.
\\\\
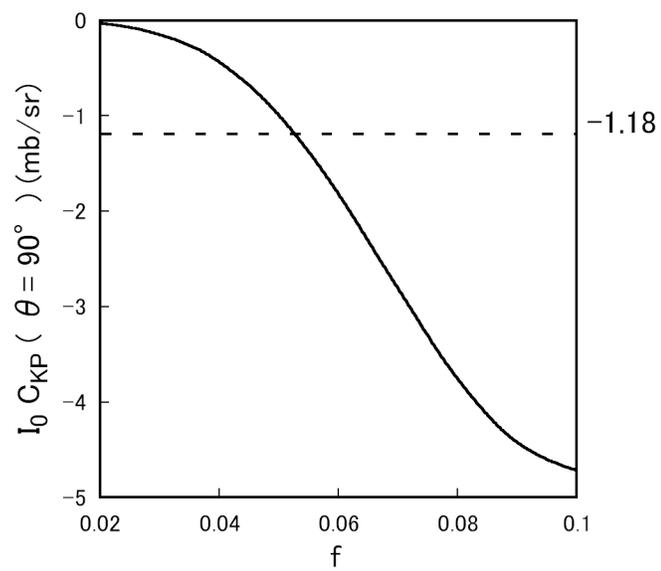
Figure 2: $I_{\rm 0}C_{KP}$ as a function of the pseudovector coupling constant $f$
at the center of mass scattering angle $\theta=90^\circ$ and the laboratory energy of 310 MeV. 
The experimental value -1.18 is from $C_{KP}$ data of 400 MeV in ref. $\cite{Engels}$
and $I_0$ value at 312 MeV and $\theta=30.67^\circ$ in ref. $\cite{Besset}$.
\\\\
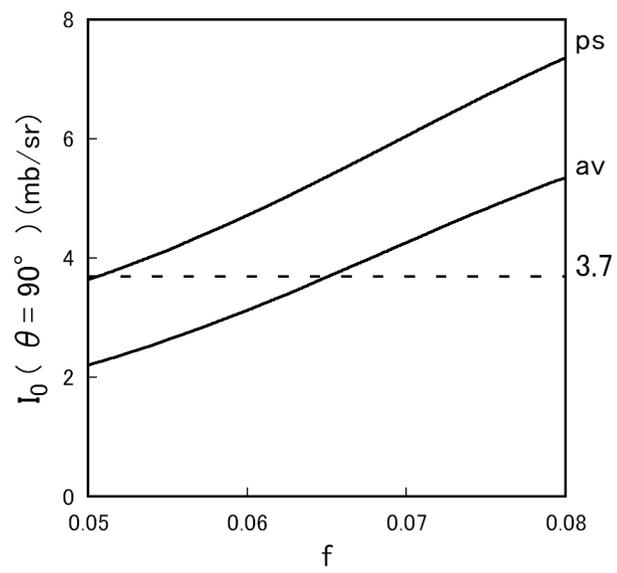
Figure 3: $I_{\rm 0}$ as a function of the pseudovector coupling constant $f$
at the center of mass scattering angle $\theta=90^\circ$ and the laboratory energy of 310 MeV. 
The $av$ and $ps$ denote same as that in Fig. 1.
The experimental value 3.7 is from the data at 312 MeV and $\theta=30.67^\circ$ in ref. $\cite{Besset}$.
\\\\
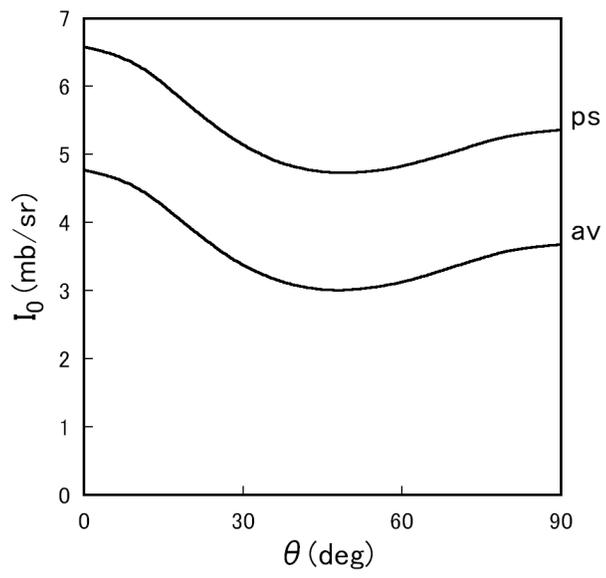
Figure 4: The differential cross section $I_{\rm 0}$ as a function of the center of mass scattering angle $\theta$
at the laboratory energy of 310 MeV. 
The $av$ and $ps$ denote same as that in Fig. 1.
\\\\
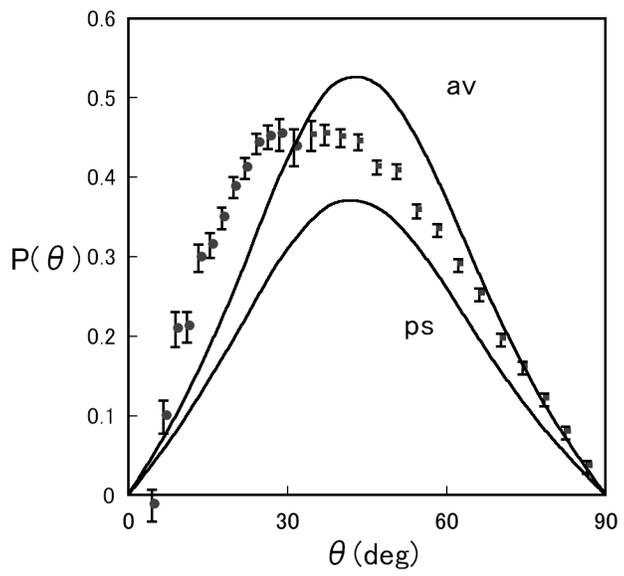
Figure 5: The polarization $P(\theta)$ as a function of the center of mass scattering angle $\theta$
at the laboratory energy of 400 MeV. 
The $av$ and $ps$ denote same as that in Fig. 1.
The experimental data are from the measurements at 392 MeV laboratory energy in ref. $\cite{Besset}$
and 396 MeV in ref. $\cite{Besset2}$.
\\\\
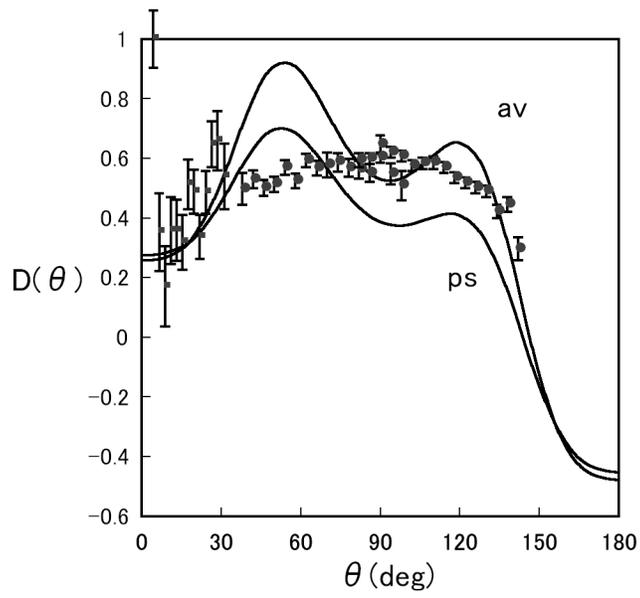
Figure 6: The depolarization $D(\theta)$ as a function of the center of mass scattering angle $\theta$
at the laboratory energy of 400 MeV. 
The $av$ and $ps$ denote same as that in Fig. 1.
The experimental data are from the measurements at 392 MeV laboratory energy in ref. $\cite{Besset}$
and 366 MeV and 398 MeV in ref. $\cite{Onel}$.
\\\\
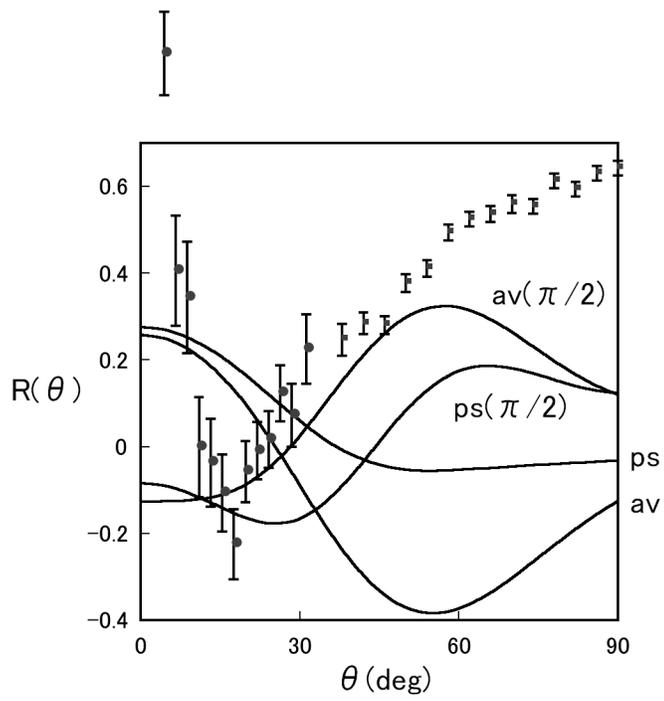
Figure 7: The spin rotation parameter $R(\theta)$ as a function of the center of mass scattering angle $\theta$
at the laboratory energy of 400 MeV. 
The $av$ and $ps$ denote same as that in Fig. 1.
The $av(\pi/\rm 2)$ and $ps(\pi/\rm 2)$ denote the axial-vector and the pseudoscalar components for spin singlet part respectively with the resonance effect.
The experimental data are from the measurements at 392 MeV laboratory energy in ref. $\cite{Besset}$
and $D_{\omega0s0}$ at 398 MeV in ref. $\cite{Onel}$.
\\\\\\
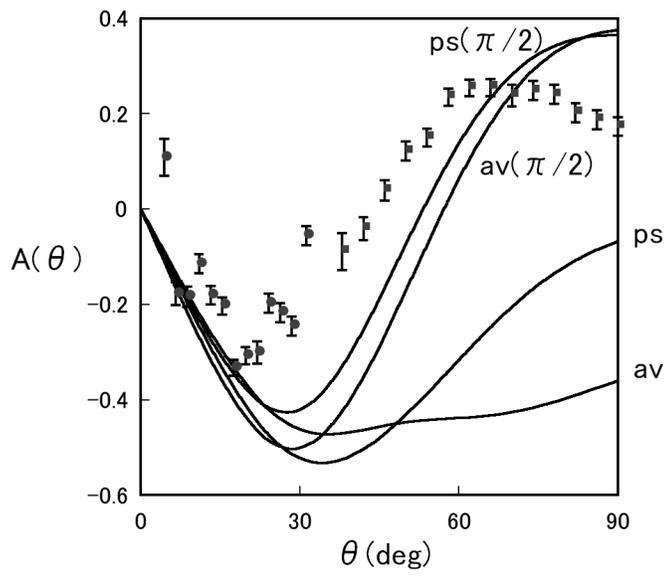
Figure 8: The spin rotation parameter $A(\theta)$ as a function of the center of mass scattering angle $\theta$
at the laboratory energy of 400 MeV. 
The $av$ and $ps$ denote same as that in Fig. 1.
The $av(\pi/\rm 2)$ and $ps(\pi/\rm 2)$ denote same as that in Fig. 7.
The experimental data are from the measurements at 392 MeV laboratory energy in ref. $\cite{Besset}$
and $D_{\omega0k0}$ at 398 MeV in ref. $\cite{Onel}$.
\end{document}